\documentclass[11pt,a4paper]{article}
\pdfoutput=1
\usepackage{jheppub}
\usepackage{rotating}
\usepackage{array}
\usepackage{amsmath}
\usepackage[normalem]{ulem}
\usepackage{slashed}
\usepackage{units}
\usepackage{enumitem}

\newlength\mylabelwd
\def\bm#1{\boldsymbol{#1}}
\def\BRcLSP{{\rm Br}}
\def\sigmastoppair{\sigma_{\tilde t_1 \tilde t_1^\ast}}

\preprint{CERN-TH-2016-074 \\ \phantom{xxx} \hfill  PSI-PR-16-03}

\title{Stop searches in flavourful supersymmetry}

\author[a,d]{Andreas Crivellin,}
\author[b,d]{Ulrich Haisch}
\author[c]{and Lewis~C.~Tunstall}

\affiliation[a]{Paul Scherrer Institut, CH-5232 Villigen PSI, Switzerland}
\affiliation[b]{Rudolf Peierls Centre for Theoretical Physics, University of Oxford, \\ 1 Keble Road, Oxford OX1 3NP, United Kingdom}
\affiliation[c]{Albert Einstein Center for Fundamental Physics,
Institute for Theoretical Physics,\\ University of Bern, Sidlerstrasse 5, 
CH-3012 Bern, Switzerland}
\affiliation[d]{CERN, Theory Division, CH-1211 Geneva 23, Switzerland}

\emailAdd{andreas.crivellin@cern.ch}
\emailAdd{ulrich.haisch@physics.ox.ac.uk}
\emailAdd{tunstall@itp.unibe.ch}

\abstract{Natural realisations of supersymmetry require light stops ${\tilde t}_1$, making them a prime target of LHC searches for physics beyond the Standard Model. Depending on the kinematic region, the main search channels are ${\tilde t_1}\to t \tilde \chi^0_1$, ${\tilde t_1}\to W b \tilde \chi^0_1$ and ${\tilde t_1}\to c \tilde \chi^0_1$. We first examine the interplay of these decay modes with ${\tilde c_1}\to c \tilde \chi^0_1$ in a model-independent fashion, revealing the existence of large regions in parameter space which are excluded for any ${\tilde t_1}\to c \tilde \chi^0_1$ branching ratio. This effect is then illustrated for scenarios with stop-scharm mixing in the right-handed sector, where it has previously been observed that the stop mass limits can be significantly weakened for large mixing. Our analysis shows that once the~LHC bounds from ${\tilde c_1}\to c \tilde \chi^0_1$ searches are taken into account, non-zero stop-scharm mixing leads only to a modest increase in the allowed regions of parameter space, with large areas excluded for arbitrary mixing angles.}

\begin{document}
\maketitle
\flushbottom

\section{Introduction}
\label{sec:introduction}

A key feature of supersymmetric extensions of the Standard Model (SM) is the fact that radiative corrections to the Higgs potential can induce electroweak symmetry breaking in a technically natural fashion. Since top quarks and top squarks dominate the radiative corrections, naturalness requires their masses to be of similar magnitudes to ensure a sufficient cancellation of quadratic divergences.  Apart from the gluino, Higgsinos and the left-handed bottom squark, the rest of the superpartners are less important for naturalness, and may well have masses above the reach of the LHC~\cite{Dimopoulos:1995mi,Pomarol:1995xc,Cohen:1996vb, Kitano:2006gv,Brust:2011tb,Papucci:2011wy}. A spectrum with the above hierarchy is a typical starting point for phenomenological analyses in supersymmetry (SUSY).   

Although light stops are required for naturalness, they can reintroduce fine-tuning in minimal SUSY because the Higgs is typically predicted to be light. For instance, to accommodate a Higgs mass of $125 \, {\rm  GeV}$ in the Minimal Supersymmetric SM (MSSM), the stop masses must be around $1 \, {\rm TeV}$, at the cost of tuning at the percent level or worse.  Reconciling these two features, light stops for naturalness and heavier stops for the Higgs mass, constitutes the  ``little hierarchy problem''. However, in contrast to naturalness, the little hierarchy problem is model dependent and tightly bound to the MSSM. SUSY models that can generate a sufficiently heavy Higgs with improved naturalness include scenarios with non-decoupling $D$-terms~\cite{Batra:2003nj} and the next-to-minimal supersymmetric SM with special parameter choices~\cite{Hall:2011aa}.

Naturalness considerations aside, there are additional reasons to expect light stops if SUSY is realised in nature. For instance, the renormalisation group evolution from a high scale with universal squark masses typically drives the masses of the third generation squarks to small values~\cite{Martin:1997ns}. Light stops also help accommodate the observed dark matter relic density~\cite{Boehm:1999bj,Ellis:2001nx} and are an essential ingredient in realising baryogenesis~\cite{Carena:1996wj,Espinosa:1996qw,Delepine:1996vn}. 

Experimentally, the bounds on the lightest stop mass~$m_{\tilde{t}_1}$ are much weaker than the limits on the other coloured superpartners,~i.e.~the squarks of the first two generations and the gluino~\cite{ATLASsummary,CMSsummary}. There are three main kinematic regions where different channels are used to search for stops, namely 
\begin{enumerate}[label=R\arabic*), align=left, labelindent=\parindent, leftmargin=\dimexpr\mylabelwd+\labelindent+\labelsep\relax, itemindent=*]
\item $m_{\tilde t_1}-m_{\tilde\chi^0_1}>m_t$:  $\tilde t_1\to t\tilde\chi^0_1$ ,
\item $m_W+m_b<m_{\tilde t_1}-m_{\tilde\chi^0_1}<m_t$: $\tilde t_1\to W b\tilde\chi^0_1$ ,
\item $m_c<m_{\tilde t_1}-m_{\tilde\chi^0_1}<m_W+m_b$: $\tilde t_1\to c\tilde\chi^0_1$ and  $\tilde t_1\to bff'\tilde\chi^0_1$ .
\end{enumerate}
Here $m_{\tilde\chi^0_1}$ denotes the mass of the lightest neutralino, constituting the lightest superpartner~(LSP), while  $m_W$, $m_b$ and $m_c$ are the mass of the $W$ boson, the bottom quark and the charm quark, respectively.  

In each region, the results from the ATLAS and CMS searches are interpreted in the context of simplified models, where the branching ratio for each decay mode is fixed to~100\% and flavour violation is assumed to be absent. Under these assumptions, the resulting limits on $m_{\tilde t_1}$ in the region R1 are strong, reaching up to stop masses close to~$800 \, {\rm GeV}$~\cite{Aad:2014qaa,Aad:2014bva,Aad:2014kra,Chatrchyan:2013xna,CMS:2013nia,CMS:2014wsa,ATLAS-CONF-2016-007,ATLAS-CONF-2016-009,CMS-PAS-SUS-16-002,CMS-PAS-SUS-16-004}. It has been observed~\cite{Bartl:2012tx,Blanke:2013zxo,Agrawal:2013kha,Blanke:2015ulx}, however,  that these limits can be weakened if non-minimal sources of flavour violation are present. This occurs because flavour-violating effects enhance the decay width for $\tilde t_1\to c\tilde\chi^0_1$, and thereby reduce the branching ratio for $\tilde t_1\to t\tilde\chi^0_1$ from unity.   On the other hand, if the decay width of $\tilde t_1\to c\tilde\chi^0_1$ becomes large, the limits from direct $\tilde c_1$ pair production and subsequent scharm decay $\tilde c_1\to c\tilde\chi^0_1$~\cite{Aad:2015gna} become relevant, which  apply to~$\tilde t_1\to c\tilde\chi^0_1$ as well once the branching ratio is large. In the second region R2, the situation is similar. The limits on $m_{\tilde t_1}$ reach only up to around~$300 \, {\rm GeV}$~\cite{Chatrchyan:2013xna,Aad:2014kra,Aad:2014qaa} and the three-body decay $\tilde t_1\to W b\tilde\chi^0_1$ is suppressed by phase space, so that $\tilde t_1\to c\tilde\chi^0_1$ can compete for relatively small off-diagonal elements in the squark mass matrix~\cite{Grober:2015fia}. Again, once stop-scharm mixing and therefore the decay width for $\tilde t_1\to c\tilde\chi^0_1$ is sizeable,  searches for charm signatures~\cite{Aad:2014nra} can become relevant. Finally, in the third region R3, $\tilde t_1\to c\tilde\chi^0_1$ is typically the dominant decay mode and the four-body decay~$\tilde t_1\to b ff^\prime \tilde\chi^0_1$~\cite{Aad:2014kra,Aad:2014nra} can only compete for scenarios resembling Minimal Flavour Violation (MFV)~\cite{Grober:2014aha}. 

The purpose of this article is to examine the complementarity of $\tilde c_1\to c\tilde\chi^0_1$ searches with the standard channels $\tilde t_1\to W b\tilde\chi^0_1$ and  $\tilde t_1\to c\tilde\chi^0_1$ in the presence of non-minimal sources of flavour violation. In Section~\ref{sec:R1}, we introduce the basic ideas behind our combination procedure and apply it to set model-independent limits on $m_{\tilde{t}_1}$,  $m_{\tilde\chi^0_1}$ and the branching ratio of $\tilde t_1\to c\tilde\chi^0_1$ in the kinematic region R1 using ATLAS Run~I data. The very same exercise is performed in Section~\ref{sec:R2} for the region R2. Focusing on flavour mixing in the right-handed up-squark sector, which is largely unconstrained by quark flavour observables, we then quantify in  Section~\ref{sec:RR} the interplay between the different search strategies. As we are interested in non-MFV scenarios in this article, we use the ATLAS bounds for $\tilde t_1\to c\tilde\chi^0_1$ directly in region R3. This allows us to provide interesting exclusions in large parts of the entire $m_{\tilde{t}_1}$--$\hspace{0.75mm} m_{\tilde\chi^0_1}$ plane. Our conclusions and an outlook are presented in Section~\ref{sec:conclusions}. In order to make our article self-contained, Appendix~\ref{app:A} provides details on the Monte Carlo (MC) simulations that were used to obtain the numerical results presented in our work.

\section{Stop search combination for $\bm{m_{\tilde t_1}-m_{\tilde\chi^0_1}>m_t}$}
\label{sec:R1}

We begin our numerical analysis in the kinematic region R1. In this region the two-body decay $\tilde t_1 \to t \tilde \chi_1^0$ dominates unless non-minimal sources of flavour violation in the up-squark sector are present that lead to an appreciable rate for $\tilde t_1 \to c \tilde \chi_1^0$. As illustrated in Figure~\ref{fig:diagrams1}, in such cases one faces three different decay configurations: one that involves two top quarks~(configuration~$1$), one with an intermediate top and a charm quark~(configuration~$2$), and finally one with two charm quarks~(configuration~$3$). Since the final state contains two~LSPs in all configurations, the visible decay products will be augmented by large amounts of missing transverse momentum ($E_{T, \rm miss})$.

In order to find combined model-independent limits on $m_{\tilde{t}_1}$,  $m_{\tilde\chi^0_1}$ and the branching ratio ${\rm Br} \left (\tilde t_1\to c\tilde\chi^0_1 \right)$   in the region R1, we employ three different ATLAS searches that are all based on around $20 \, {\rm fb}^{-1}$ of $\sqrt{s} = 8 \, {\rm TeV}$ data. Specifically, these are 
\begin{enumerate}[align=left, labelindent=\parindent, leftmargin=\dimexpr\mylabelwd+\labelindent+\labelsep\relax, itemindent=*]

\item[$a$)] $2~c\text{-tags} + E_{T, \rm miss}$ \cite{Aad:2015gna}: This ATLAS search is originally designed for the decay configuration 3 in Figure~\ref{fig:diagrams1}. In order to maximise the sensitivity of this search, three distinct signal regions~(SRs) called {\tt mct150},  {\tt mct200}  and   {\tt mct250} are defined. In all SRs, events have   to have a reconstructed primary vertex consistent with the beam positions and to meet basic quality criteria. Furthermore, events are required to contain no residual electron or muon candidate and at least two jets with radius $R = 0.4$ and $p_T > 130, 100 \, {\rm GeV}$ and $|\eta|< 2.5$. The multijet background contribution with large $E_{T, \rm miss}$ is suppressed by requiring a minimum azimuthal separation $|\Delta \phi (\vec{p}_{T,j_{1,2,3}}, \vec{p}_{T,\rm miss})|  > 0.4$ between any of the three leading jets and the $E_{T, \rm miss}$ direction $\vec{p}_{T,\rm miss}$. The third jet is exempted from this angular requirement, if it has $p_T < 50 \, {\rm GeV}$, $|\eta| < 2.4$ and less than half of the sum of its track $p_T$ is associated with tracks matched to the primary vertex. The two highest-$p_T$ jets are required to be identified as arising from a charm quark~($c$-tagged). The algorithm used in the ATLAS analysis achieves a $c$-tagging efficiency of 20\% with a $b$-jet and light-jet rejection fraction of 8 and 200~(medium operating point)~\cite{ATL-PHYS-PUB-2015-001}.  The $E_{T, \rm miss}$ selections are $E_{T, \rm miss} > 150 \, {\rm GeV}$ and $E_{T, \rm miss}/\sum_{i=1,2} |p_{T, j_i}| > 1/3$. To further discriminate between signal and background  the invariant mass of the two $c$-tagged jets has to satisfy $m_{c\bar c} > 200 \, {\rm GeV}$ and a selection based on the boost-corrected contransverse mass $m_{\rm CT}$~\cite{Polesello:2009rn} is employed.  Depending on the SR, $m_{\rm CT} > 150 \, {\rm GeV}$, $m_{\rm CT} > 200 \, {\rm GeV}$ or $m_{\rm CT} > 250 \, {\rm GeV}$ is required.

\begin{figure}
\begin{center}
\includegraphics[height=0.15\textheight]{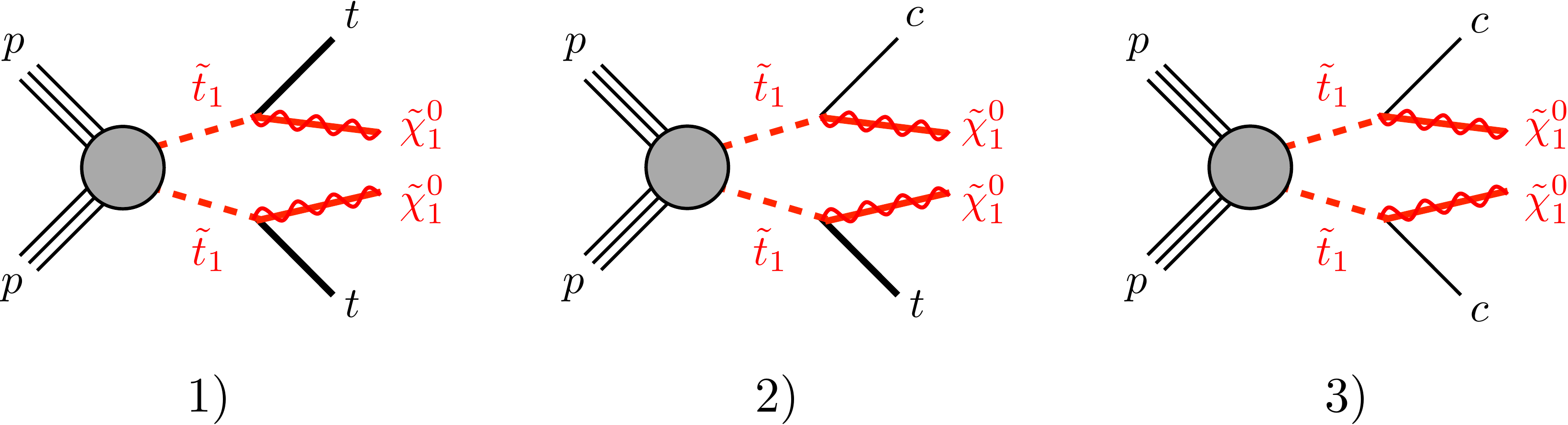} 
\end{center}
\vspace{-6mm}
\caption{\label{fig:diagrams1} The three different decay configurations relevant for the combination of different stop channels  in the kinematic region R1.}
\end{figure}
 
\item[$b$)] $1~\text{lepton} + 4~\text{jets} + 1~b\text{-tag} + E_{T, \rm miss}$ \cite{Aad:2014kra}:  In its original form, this ATLAS search has been tailored for the decay configuration 1 in Figure~\ref{fig:diagrams1} with one top quark decaying hadronically and the other one leptonically. It implements four  SRs that target different regions in the  $m_{\tilde{t}_1}$--$\hspace{0.75mm} m_{\tilde\chi^0_1}$ plane and implement different analysis strategies. In our case it turns out that only the SRs called {\tt  tN\_diag} and~{\tt  tN\_med} are relevant in the combination.   The following preselection  criteria  are common to the two SRs. Events are required to have a reconstructed primary vertex, $E_{T,\rm miss} > 100 \, {\rm GeV}$, exactly one isolated lepton with $p_T > 25 \, {\rm GeV}$ and at least four $R = 0.4$ jets with $p_T > 25 \, {\rm GeV}$.  Events that do not pass certain data quality requirements are rejected. In  the SR {\tt  tN\_diag}, the cuts on the three hardest jets are $p_T > 60, 60, 40 \, {\rm GeV}$ and $|\eta | < 2.5$. The two leading jets have to satisfy $|\Delta\phi (\vec{p}_{T,j_{1,2}}, \vec{p}_{T, \rm miss})| > 0.8$ and at least one jet has to be identified as a bottom-quark jet ($b$-tagged), assuming an average tagging efficiency of 70\%~\cite{ATLAS-CONF-2011-089,ATLAS-CONF-2011-102}. In addition, we require in our analysis $E_{T,\rm miss} > 150 \, {\rm GeV}$, $E_{T,\rm miss}/\sqrt{H_T} > 5 \, {\rm GeV}^{1/2}$, $m_T > 140 \, {\rm GeV}$, $m_{\text{had-top}} \in [130, 205] \, {\rm GeV}$ and impose a veto on loose $\tau$ leptons. Here  $H_T$ is defined as the scalar $p_T$ of the four hardest jets in the event, $m_T$ denotes the transverse mass constructed from the lepton transverse momentum and $\vec{p}_{T,\rm miss}$, while $m_{\text{had-top}}$ represents the hadronic top mass.  The selection requirements in~{\tt  tN\_med} that differ from to that of {\tt  tN\_diag}  are $p_T > 80, 60, 40 \, {\rm GeV}$ for the three leading jets,  $|\Delta\phi (\vec{p}_{T,j_{2}}, \vec{p}_{T, \rm miss})| > 0.8$, $E_{T,\rm miss} > 200 \, {\rm GeV}$ and $m_{\text{had-top}} \in [130, 195] \, {\rm GeV}$. A cut on $E_{T,\rm miss}/\sqrt{H_T}$ and a $\tau$-veto is not imposed, but  $H_{T, \rm miss}^{\rm sig} > 12.5$ and $am_{T2} > 170 \, {\rm GeV}$ is required. Here $H_{T, \rm miss}^{\rm sig}$ is an object-based missing transverse momentum that is normalised by the per-event resolution of the jets~\cite{Aad:2014kra} and $am_{T2}$ is an asymmetric variant of the generalised transverse mass~\cite{Cheng:2008hk,Barr:2009jv,Konar:2009qr,Bai:2012gs}. 

\item[$c$)] $6~\text{jets} + 2~b\text{-tags} + E_{T, \rm miss}$ \cite{Aad:2014bva}:  This ATLAS search aims to provide the best sensitivity for the decay configuration 1 in Figure~\ref{fig:diagrams1} with both top quarks decaying hadronically. In our analysis, we consider only the SR {\tt A1} and the SR {\tt A2} out of the possible nine~SRs.  All events that pass certain quality requirements and do not contain a reconstructed  electron or muon with $p_T > 10 \, {\rm GeV}$ are subjected to the following common selection criteria. They have to have at least six~$R = 0.4$ jets with $p_T > 80, 80, 35, 35, 35, 35 \, {\rm GeV}$ and $|\eta| < 2.8$, and out of these jets, two or more have to be $b$-tagged (70\% efficiency). The number of events with mismeasured $E_{T, \rm miss}$ is reduced by requiring $|\Delta\phi (\vec{p}_{T,j_{1,2,3}}, \vec{p}_{T, \rm miss})| > \pi/5$ and $|\Delta\phi ( \vec{p}_{T, \rm miss},  \vec{p}_{T, \rm miss}^{\rm \; track})| < \pi/3$, where $\vec{p}_{T, \rm miss}^{\rm \; track}$ denotes the missing transverse momentum direction determined from the calorimeter system. To further sculpt the signal, the transverse mass calculated from the $b$-tagged jet closest in the azimuthal angle~$\phi$ to $\vec{p}_{T, \rm miss}$ has to satisfy $m_T^{b,\rm min} > 175 \, {\rm GeV}$, the mass cuts $m_{bjj}^0 < 225 \, {\rm GeV}$ and  $m_{bjj}^1 < 250 \, {\rm GeV}$ on the first and second top candidate~\cite{Aad:2014bva} are imposed and loose~$\tau$~leptons are vetoed. The SRs~{\tt A1} and {\tt A2} only differ in the imposed $E_{T,\rm miss}$ selection. In the former case, events with $E_{T,\rm miss} > 150 \, {\rm GeV}$ suffice,  while in the latter case the stronger requirement $E_{T,\rm miss} > 250 \, {\rm GeV}$ is imposed.  

\end{enumerate}

To combine the searches $a$, $b$ and $c$, we work in the narrow-width approximation and assume that only the decay modes $\tilde t_1 \to t \tilde \chi_1^0$ and~$\tilde t_1 \to c \tilde \chi_1^0$ are relevant, so that ${\rm Br} \left (\tilde t_1 \to t \tilde \chi_1^0 \right ) = 1- {\rm Br} \left (\tilde t_1 \to c \tilde \chi_1^0 \right )$. Both assumptions are satisfied in the kinematic region~R1. Using the shorthand notations $\BRcLSP = {\rm Br} \left (\tilde t_1 \to c \tilde \chi_1^0 \right )$ and  $\sigmastoppair = \sigma \left (pp \to \tilde t_1 \tilde t_1^\ast \right)$, the fiducial cross sections $(\sigma_{\rm fid})_s$ corresponding to the three different ATLAS searches can then be written in the following way 
\begin{equation} \label{eq:fidxsecR1}
\begin{split}
\left ( \sigma_{\rm fid} \right )_a & = \left \{  \big (1 - \BRcLSP  \big )^2   \hspace{0.5mm} \epsilon_{1a}  + 2  \hspace{0.25mm} \BRcLSP  \hspace{0.5mm}  \big  (1 - \BRcLSP \big )  \hspace{0.5mm} \epsilon_{2a} + \BRcLSP^2  \right \} \sigmastoppair  \,, \\[2mm]
\left ( \sigma_{\rm fid} \right )_b & = \left \{  \big ( 1- \BRcLSP  \big )^2  + 2  \hspace{0.25mm} \BRcLSP \hspace{0.5mm} \big  (1 - \BRcLSP  \big )  \hspace{0.5mm} \epsilon_{2b} + \BRcLSP ^2   \hspace{1mm} \epsilon_{3b}   \right \} \sigmastoppair \,, \\[2mm]
\left ( \sigma_{\rm fid} \right )_c & = \left \{  \big ( 1- \BRcLSP  \big )^2  + 2  \hspace{0.25mm} \BRcLSP \hspace{0.5mm} \big  (1 - \BRcLSP  \big )  \hspace{0.5mm} \epsilon_{2c} + \BRcLSP^2   \hspace{1mm} \epsilon_{3c}   \right \} \sigmastoppair \,. 
\end{split}
\end{equation} 
Here $\epsilon_{ts}$ denotes the efficiency with which the decay configuration $t=1,2,3$ (see Figure~\ref{fig:diagrams1}) is detected by the search $s = a,b,c$. 

\begin{figure}
\begin{center}
\includegraphics[height=0.535\textheight]{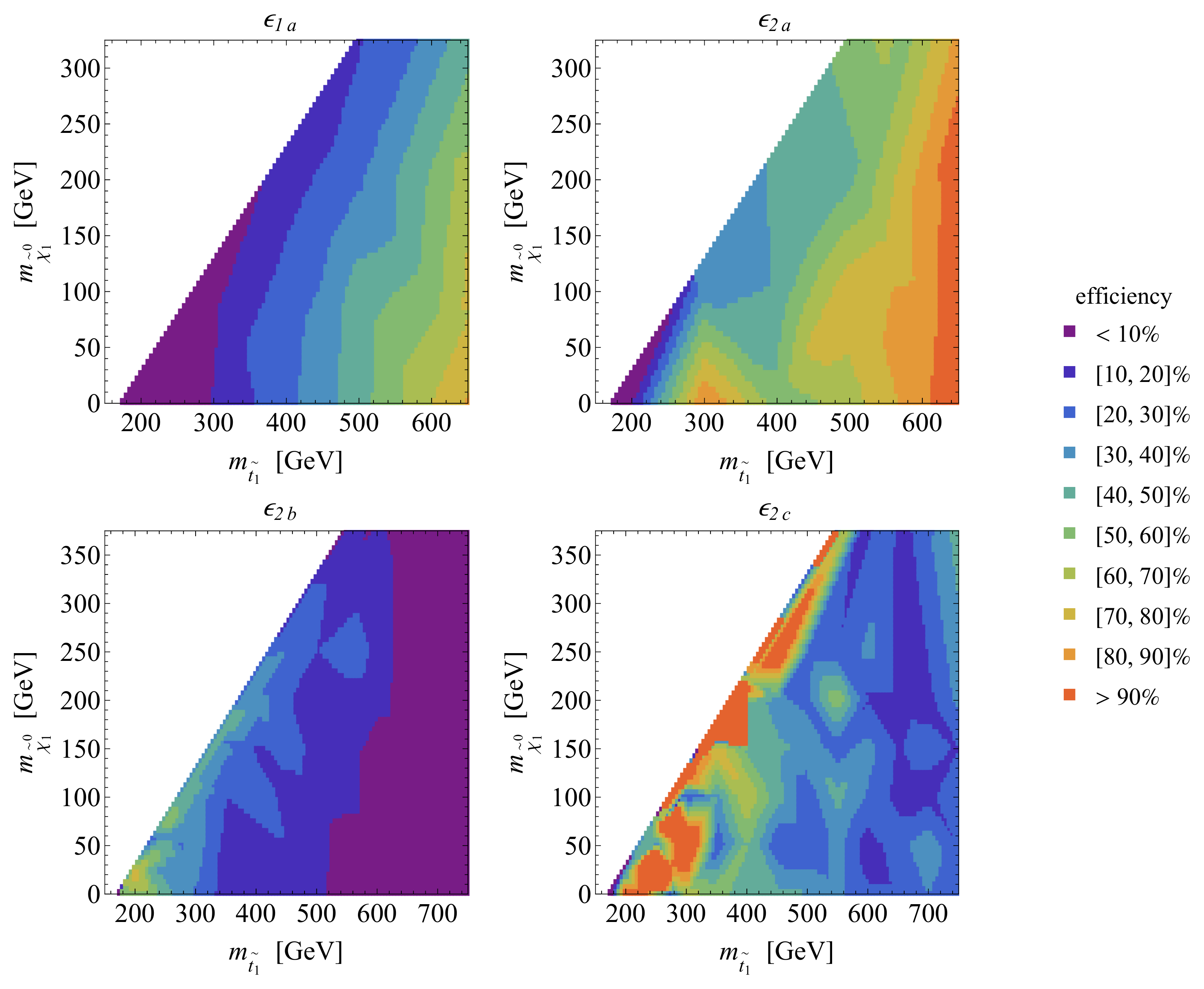} 
\end{center}
\vspace{-6mm}
\caption{\label{fig:efficiencies1} Efficiency maps relevant for the combination of different stop channels  in the kinematic region R1. Only the non-trivial efficiencies $\epsilon_{1a}$~(upper left panel),  $\epsilon_{2a}$~(upper right panel),  $\epsilon_{2b}$~(lower left panel) and  $\epsilon_{2c}$~(lower right panel) are shown.}
\end{figure}

The efficiency maps relevant  for the combination of  the different stop channels in the  region R1 are displayed  in the four panels of Figure~\ref{fig:efficiencies1}. They have been obtained by means of the MC simulations described in Appendix~\ref{app:A}. From the plots it is evident that the efficiencies $\epsilon_{ts}$ are not flat, but depend rather sensitively on $m_{\tilde{t}_1}$ and $\hspace{0.75mm} m_{\tilde\chi^0_1}$. This behaviour is expected because changing the mass of the lightest stop and the LSP will modify the kinematic distributions of the final-state particles, which in turn leads to different signal acceptances in the various SRs. In fact, a qualitative understanding of the obtained efficiencies  is possible by studying the cutflow of the analysis $a$, $b$ and $c$ for the different signal configurations $1$, $2$ and $3$. We start by discussing the efficiencies $\epsilon_{1a}$  and $\epsilon_{2a}$ shown in the upper left and upper right panel of Figure~\ref{fig:efficiencies1}, respectively. The first observation is that in most parts of the $m_{\tilde{t}_1}$--$\hspace{0.75mm} m_{\tilde\chi^0_1}$ plane, the efficiency~$\epsilon_{1a}$ is smaller than $\epsilon_{2a}$. This is readily understood by recalling from Figure~\ref{fig:diagrams1} that configuration 1 (2) leads to a final state with two bottom quarks~(one bottom quark). In the former  case, two $b$ quarks have to be misidentified as $c$-jets in order to produce an event in the SRs of search $a$, while in the latter case one mis-tag is sufficient. Another feature that is evident from the plots is that the efficiencies~$\epsilon_{1a}$  and $\epsilon_{2a}$ both decrease if one approaches the kinematic boundary  of region R1. This is due to the fact that  the decay chain $\tilde t_1 \to t \tilde \chi_1^0 \to W^+ b  \chi_1^0$ and its conjugate will not give rise to significant $E_{T,\rm miss}$ if the mass difference $m_{\tilde t_1}-m_{\tilde\chi^0_1}$ is close to~$m_t$, and as a result the corresponding event is less likely to pass the $E_{T,\rm miss}$ requirements that are imposed in the scalar charm search $a$. 

\begin{figure}
\begin{center}
\includegraphics[height=0.27\textheight]{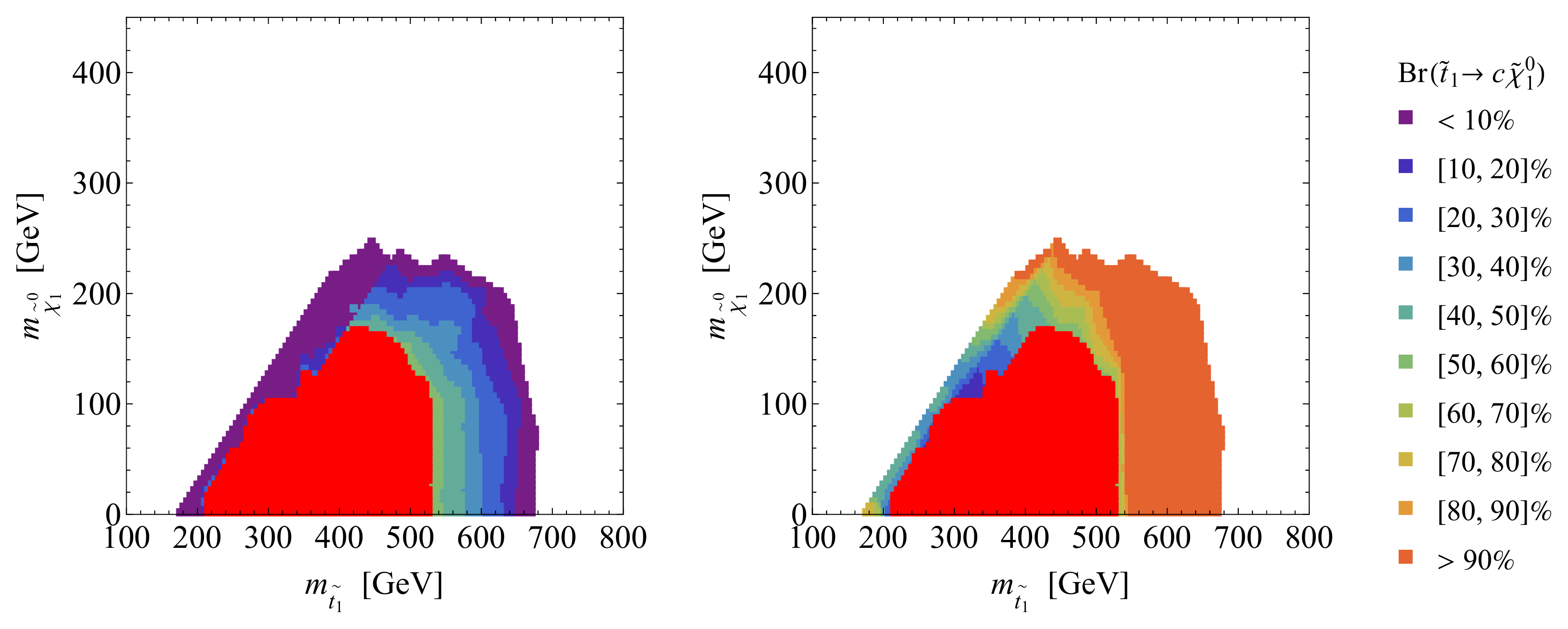} 
\end{center}
\vspace{-8mm}
\caption{\label{fig:exclusionsR1} Lower (left panel) and upper (right panel) model-independent  limit on ${\rm Br} \left (\tilde t_1 \to c \tilde \chi_1^0 \right)$ in the  part of the $m_{\tilde{t}_1}$--$\hspace{0.75mm} m_{\tilde\chi^0_1}$ plane corresponding to the kinematic region R1. The regions coloured red are excluded at 95\%~CL for any value of the $\tilde t_1 \to c \tilde \chi_1^0$ branching ratio.}
\end{figure}

Simple qualitative explanations of the efficiency maps $\epsilon_{2b}$ and $\epsilon_{2c}$ presented in the lower left and lower right panel of Figure~\ref{fig:efficiencies1} can also be given.  In the case of $\epsilon_{2b}$, the requirement to have an isolated lepton strongly suppresses the acceptance for final states arising from the configuration 2, which involves both a $\tilde t_1 \to t \tilde \chi_1^0 \to W^+ b \chi_1^0$ and a  $\tilde t_1^\ast \to \bar c \tilde \chi_1^0$ decay or the combination of charge-conjugated processes. For these decays, the fact that  a lepton  with~$p_T > 25 \, {\rm GeV}$ can only arise from a leptonic decay of a $W$ boson also explains the finding that $\epsilon_{3b} \simeq 0$ in the whole kinematic region R1. That $\epsilon_{2b}$ increases when approaching the kinematic boundary $m_{\tilde t_1} - m_{\tilde \chi_1^0} = m_t$  has to do with the fact that this efficiency is defined relative to the acceptance corresponding to a signal from the configuration 1 in the search~$b$~$\big($see (\ref{eq:fidxsecR1})$\big)$. The latter acceptance is however suppressed close to the kinematic boundary because there is only little $E_{T, \rm miss}$ available. Similar arguments hold for $\epsilon_{2c}$. In this case the requirement that events have to contain two $b$-tags plays the role that the single-lepton tag played before. In fact, events resulting from configuration 2 can  end up  in the SR, if the final-state charm quark is erroneously $b$-tagged.  The corresponding probability is non-negligible and taking into account that the relative acceptance for detecting the configuration 2 in  the search $c$ again increases for decreasing mass splitting $m_{\tilde t_1} - m_{\tilde \chi_1^0}$, one obtains numerically $\epsilon_{2c} \simeq 1$ for points with $m_{\tilde t_1} - m_{\tilde \chi_1^0}$ not too far from $m_t$. Since the probability to mistake two charm quarks for two $b$-jets is essentially zero,  we furthermore find that $\epsilon_{3c} \simeq 0$ for all points of interest in the $m_{\tilde{t}_1}$--$\hspace{0.75mm} m_{\tilde\chi^0_1}$ plane. 

With the efficiency maps at hand, one can then use (\ref{eq:fidxsecR1}) and combine the individual searches to obtain model-independent exclusion limits on ${\rm Br} \left ( \tilde t_1 \to c \tilde \chi_1^0 \right )$. The outcome of such an exercise is shown in~Figure~\ref{fig:exclusionsR1}. The red region in both panels is excluded at~95\%~confidence level (CL) for any value of the $\tilde t_1 \to c \tilde \chi_1^0$ branching ratio. We observe that depending on the LSP (stop) mass, values of $m_{\tilde t_1}$ up to $530 \, {\rm GeV}$  ($m_{\tilde \chi_1^0}$ up to $160 \, {\rm GeV}$) are ruled out by our combination of ATLAS Run I data. Outside the excluded region our procedure can be used to set lower and upper model-independent limits on ${\rm Br} \left ( \tilde t_1 \to c \tilde \chi_1^0 \right )$ as indicated by the coloured contours in the left and right panel of the figure. This information will be used in Section~\ref{sec:RR} to put  bounds in the $m_{\tilde{t}_1}$--$\hspace{0.75mm} m_{\tilde\chi^0_1}$ plane for the case of the MSSM with a bino-like LSP and purely right-handed  stop-scharm mixing. 

\begin{figure}
\begin{center}
\includegraphics[height=0.275\textheight]{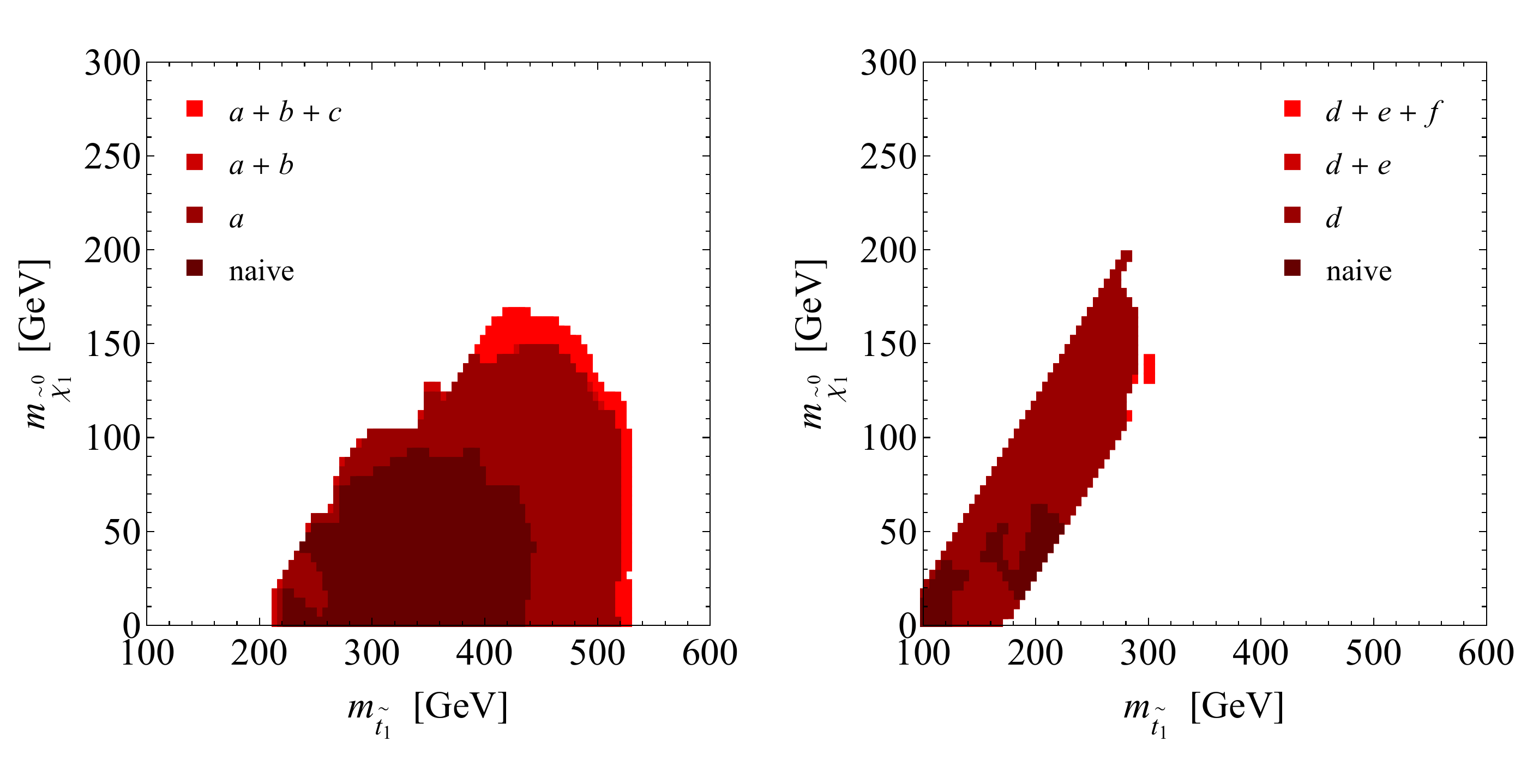} 
\end{center}
\vspace{-8mm}
\caption{\label{fig:combinations} Comparison of the impact of the different searches strategies on the 95\% CL exclusion regions in the  $m_{\tilde{t}_1}$--$\hspace{0.75mm} m_{\tilde\chi^0_1}$ plane. The left (right) panel displays the results of a naive combination and a successive inclusion of the searches $a$, $b$ and $c$ ($d$, $e$ and $f$).}
\end{figure}

It is also interesting to quantify the impact that each individual search has in the combination that leads to the final 95\% CL exclusion limit.   For the kinematic region R1, we illustrate the power of the different searches in the left panel of Figure~\ref{fig:combinations}. The naive combination corresponds to the choice~$\epsilon_{ts} = 0$ in  (\ref{eq:fidxsecR1})  and is indicated by the dark red contour in the figure. We see that a successive inclusion of the searches $a$, $b$ and $c$ enlarges the excluded area in the $m_{\tilde{t}_1}$--$\hspace{0.75mm} m_{\tilde\chi^0_1}$ plane considerably. In fact, it is evident from the three  additional red contours that the inclusion of search~$a$ has the most pronounced effect in the combination, while adding searches~$b$ and $c$ to the mix leads to either no or only a minor improvement of the exclusion limits in the $m_{\tilde{t}_1}$--$\hspace{0.75mm} m_{\tilde\chi^0_1}$ plane. This feature nicely illustrates one of the main findings of our work,~i.e.~the observation that the recent ATLAS search for~$\tilde c_1  \to c \tilde \chi_1^0$~\cite{Aad:2015gna} can be recast as  a search for $\tilde t_1  \to c \tilde \chi_1^0$, and that this procedure can be used to  set stringent bounds on $m_{\tilde t_1}$ and $ m_{\tilde\chi^0_1}$ in models with non-minimal flavour mixing in the up-squark sector. 

\section{Stop search combination  for $\bm{m_W+m_b<m_{\tilde t_1}-m_{\tilde\chi^0_1}<m_t}$}
\label{sec:R2}

\begin{figure}
\begin{center}
\includegraphics[height=0.15\textheight]{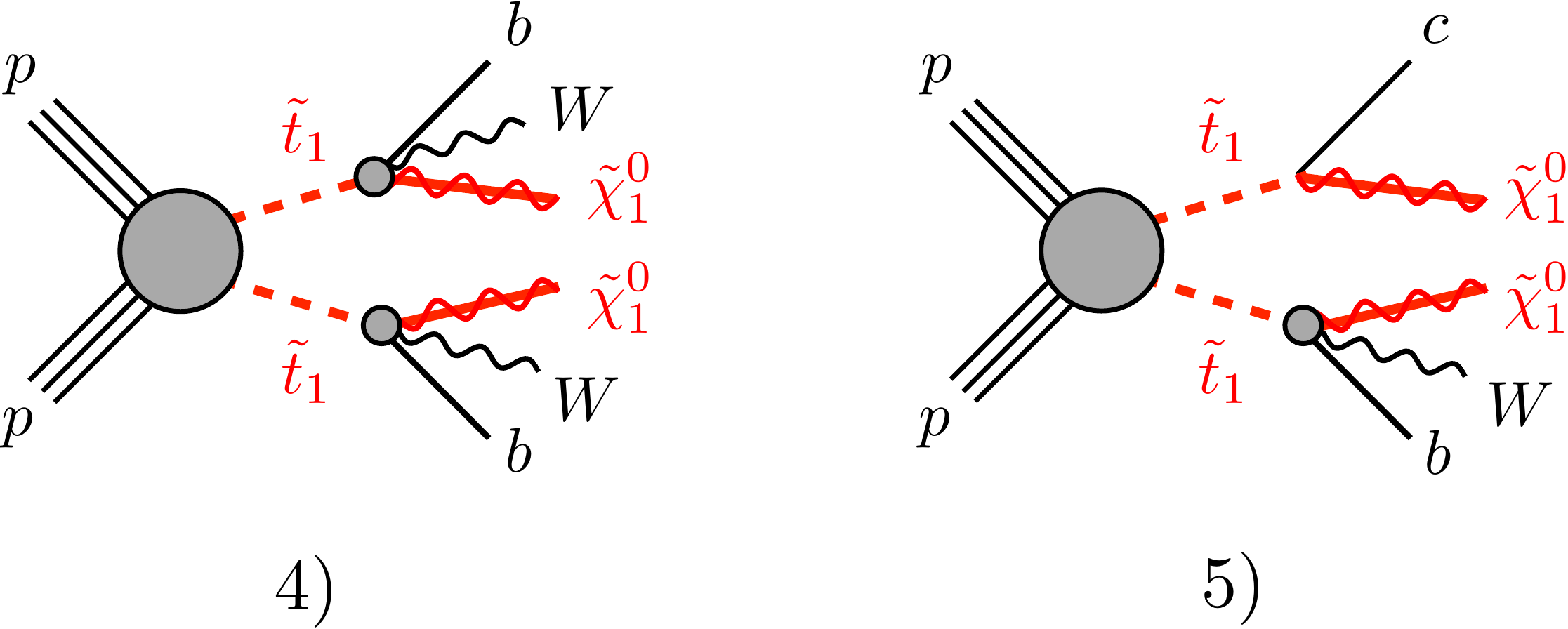} 
\end{center}
\vspace{-6mm}
\caption{\label{fig:diagrams2} The two additional  decay  configurations that are relevant for the combination of different stop searches  in the kinematic region R2.}
\end{figure}

We now turn our attention to the kinematic region R2. If  stop-scharm mixing is present, both the two-body decay $\tilde t_1 \to c \tilde \chi_1^0$ and the three-body decay $\tilde t_1 \to Wb  \tilde \chi_1^0$ can be phenomenologically relevant. As a result, the final states emerging from two stop decays can contain either two charm quarks~(configuration~$3$), two bottom quarks~(configuration~$4$) or one charm quark and one bottom quark~(configuration~$5$). The additional decay configurations with bottom quarks are depicted in Figure~\ref{fig:diagrams2}. As indicated by the small grey blobs in this figure, the $\tilde t_1 \to Wb  \tilde \chi_1^0$ transitions proceeds through an effective four-point vertex which involves the exchange of off-shell particles. In our analysis, we include for simplicity only top-quark exchange, but neglect chargino and sbottom contributions, assuming that these states are sufficiently heavy and decoupled from the spectrum.

In the kinematic region R2, we constrain the  $m_{\tilde{t}_1}$--$\hspace{0.75mm} m_{\tilde\chi^0_1}$ parameter space by again combining three different ATLAS analyses. They are all based on  $20.3 \, {\rm fb}^{-1}$ of total integrated luminosity collected at $8 \, {\rm TeV}$ centre-of-mass energy, and implement the following search strategies: 
\begin{enumerate}[align=left, labelindent=\parindent, leftmargin=\dimexpr\mylabelwd+\labelindent+\labelsep\relax, itemindent=*]

\item[d)] $4~\text{jets}  + c\text{-tags} +  E_{T, \rm miss}$ \cite{Aad:2014nra}: Originally, this  ATLAS search has been designed to gain sensitivity to the decay configuration~3 in Figure~\ref{fig:diagrams1}. In our analysis,  we consider the $c$-tagged selections {\tt C1} and {\tt C2}. The events have  to meet basic quality criteria and  are vetoed if they contain  isolated muons or isolated electrons with $p_T > 10 \, {\rm GeV}$. As a further preselection $E_{T,\rm miss} > 150 \, {\rm GeV}$ and least one $R = 0.4$ jet with $p_T   > 150 \, {\rm GeV}$ and $|\eta| < 2.5$ in the final state is required.  To be contained in the SRs, the events are required to have at least four jets with $p_T > 30 \, {\rm GeV}$, $|\eta| <2.5$ and $|\Delta \phi (\vec{p}_{T,j}, \vec{p}_{T,\rm miss})| > 0.4$. A $b$-jet veto (2.5 rejection factor) is applied to the selected jets by using a loose $c$-tag requirement (95\% efficiency)~\cite{ATL-PHYS-PUB-2015-001}. In addition, at least one of the three subleading jets has to pass  the medium $c$-tag criteria mentioned earlier in the description of search $a$. The leading jet is then required to have~$p_T > 290 \, {\rm GeV}$ and the two SRs {\tt C1} and {\tt C2} are defined with $E_{T, \rm miss} > 250 \, {\rm GeV}$ and $E_{T, \rm miss} > 350 \, {\rm GeV}$, respectively.

\item[e)] $2~\text{leptons} + \text{jets}  + E_{T, \rm miss}$  \cite{Aad:2014qaa}: This ATLAS search targets the decay configuration~4 in Figure~\ref{fig:diagrams2} with both $W$ bosons decaying leptonically.  Our analysis  includes the~SR~{\tt L90} and  {\tt L100} of this  search. After passing certain quality requirements, events are preselected if they have exactly two oppositely charged leptons (muons, electrons or one charged lepton of each flavour). At least one of these leptons must have $p_T > 25 \, {\rm GeV}$ and the invariant mass of the lepton pair has to satisfy~$m_{ll} > 20 \, {\rm GeV}$. After applying these preselections, events with $m_{ll} \in [ 71, 111 ] \, {\rm GeV}$, $|\Delta \phi ( \vec{p}_{T,j}, \vec{p}_{T,\rm miss} )| > 1$, where $j$ denotes the jet closest to the $E_{T, \rm miss}$ direction, and $|\Delta \phi ( \vec{p}_{T,llb}, \vec{p}_{T,\rm miss} )| < 1.5$ with $\vec{p}_{T,llb} = \vec{p}_{T,l_1} +  \vec{p}_{T,l_2} + \vec{p}_{T,\rm miss}$ are rejected.  The SR {\tt L90} requires $m_{{\rm T}2} > 90 \, {\rm GeV}$ but has no jet requirement, while~{\tt L100} has a tight jet selection with at least two $R = 0.4$ jets with $p_T > 100, 50 \, {\rm  GeV}$ and $|\eta| < 2.5$. Furthermore, the cut $m_{{\rm T}2} > 100 \, {\rm GeV}$ on the lepton-based stransverse mass~\cite{Lester:1999tx,Barr:2003rg} is set in {\tt L100}.

\item[f)] $1~\text{lepton} + 3~\text{jets} + b\text{-veto} + E_{T, \rm miss}$ \cite{Aad:2014kra}: At the outset, this ATLAS search is intended for the decay configuration~4 in Figure~\ref{fig:diagrams2} with one leptonic and one hadronic $W$-boson decay.  In our combination we use the {\tt bcC\_diag} SR of this analysis, which requires one central lepton with $p_T  > 25 \, {\rm GeV}$ and $|\eta | <1.2$ as well as at least three $R = 0.4$ jets with $p_T > 80,40,30 \, {\rm GeV}$ and $|\eta | < 2.5$. Out of the three jets, none are allowed to be $b$-tagged~(70\% efficiency) and the two hardest jets have to satisfy $|\Delta\phi (\vec{p}_{T,j_{1,2}}, \vec{p}_{T, \rm miss})| > 2.0, 0.8$.  The other cuts in our analysis are $E_{T,\rm miss} > 140 \, {\rm GeV}$, $E_{T,\rm miss}/\sqrt{H_T} > 5 \, {\rm GeV}^{1/2}$, $m_T > 120 \, {\rm GeV}$ and $\Delta R (\vec{p}_{T,l}, \vec{p}_{T,j_{1}}) \in [0.8, 2.4]$. The angular separation in the $\eta\hspace{0.5mm}$--$\hspace{0.25mm}\phi$ plane is defined as $\Delta R = \sqrt{(\Delta \eta)^2 + (\Delta \phi)^2}$.

\end{enumerate}

\begin{figure}
\begin{center}
\includegraphics[height=0.55\textheight]{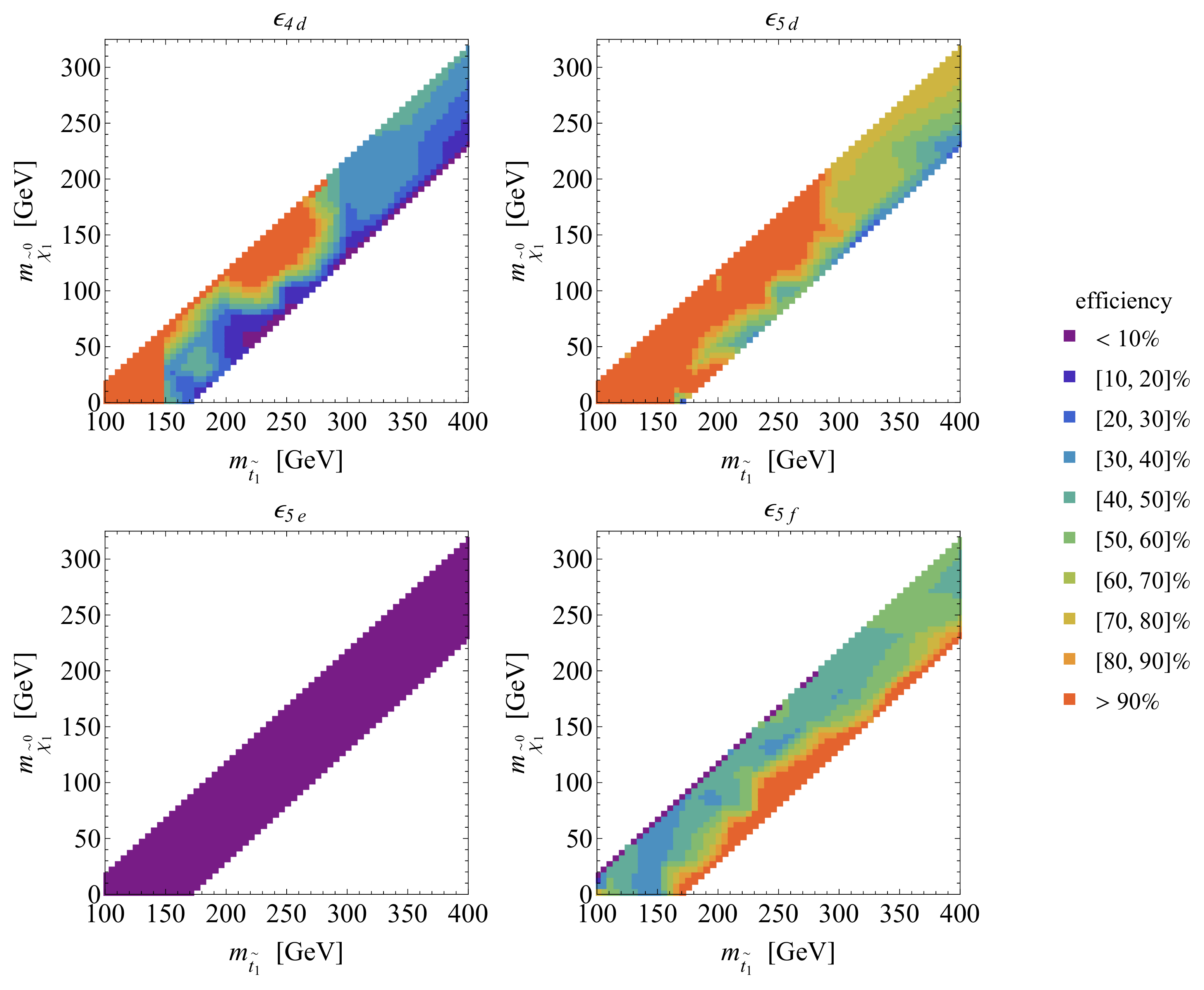} 
\end{center}
\vspace{-6mm}
\caption{\label{fig:efficiencies2} The efficiencies $\epsilon_{4d}$~(upper left panel),  $\epsilon_{5d}$~(upper right panel),  $\epsilon_{5e}$~(lower left panel) and  $\epsilon_{5f}$~(lower right panel) relevant for the combination of different stop channels  in  region R2.}
\end{figure}

Under the assumption that only the decay modes $\tilde t_1 \to W b \tilde \chi_1^0$ and~$\tilde t_1 \to c \tilde \chi_1^0$ are relevant in region R2, the searches $d$, $e$ and $f$ can then be combined by using formulas analogous to those presented in (\ref{eq:fidxsecR1}).  The corresponding efficiency maps are depicted in Figure~\ref{fig:efficiencies2}. In the upper left and right panel we show the efficiencies for detecting final states with two bottom quarks (configuration 4) or one bottom and one charm quark (configuration~5)  by means of the search~$d$. As expected the efficiency $\epsilon_{4d}$ is typically smaller than $\epsilon_{5d}$, because the search~$d$ involves a $c$-tagged selection. Another noticeable feature of the latter efficiencies is that they are enhanced close to the kinematic boundary $m_W+m_b = m_{\tilde t_1}-m_{\tilde\chi^0_1}$ for relatively light stops with $m_{\tilde t_1} \lesssim 250 \, {\rm GeV}$. This is related to the fact that, compared to~$\tilde t_1 \to c \tilde \chi_1^0$, the decay $\tilde t_1 \to Wb\tilde \chi_1^0$ produces a harder $E_{T,\rm miss}$ spectrum for masses in this region of the~$m_{\tilde{t}_1}$--$\hspace{0.75mm} m_{\tilde\chi^0_1}$ plane. As a result, events resulting from configuration~4~or~5 more easily pass the $E_{T,\rm miss}$ requirement of search $d$ than final states arising from configuration~3. 

In the lower left panel of Figure~\ref{fig:efficiencies2}, we find that the efficiency $\epsilon_{5e} \simeq 0$ in the entire R2 region, as a result of the requirement of search $e$ to have two charged leptons in each event. The very same requirement also leads to $\epsilon_{3e} \simeq 0$. The efficiency map for $\epsilon_{5f}$ is shown in the lower right panel of the latter figure. One observes that $\epsilon_{5f}$ grows  towards masses satisfying $m_{\tilde t_1}-m_{\tilde\chi^0_1} = m_t$. This feature originates mostly from the fact that for~$m_{\tilde t_1}$ and $m_{\tilde\chi^0_1}$ values close to the boundary between the regions R1 and R2, the leading jet arising from the configuration $f$ is on average harder compared to situations where the masses are close to $m_W + m_b = m_{\tilde t_1}-m_{\tilde\chi^0_1}$.  As a result of the central lepton requirement, we also find that $\epsilon_{3f} \simeq 0$ in the full R2 part of the  $m_{\tilde{t}_1}$--$\hspace{0.75mm} m_{\tilde\chi^0_1}$ plane.

\begin{figure}
\begin{center}
\includegraphics[height=0.27\textheight]{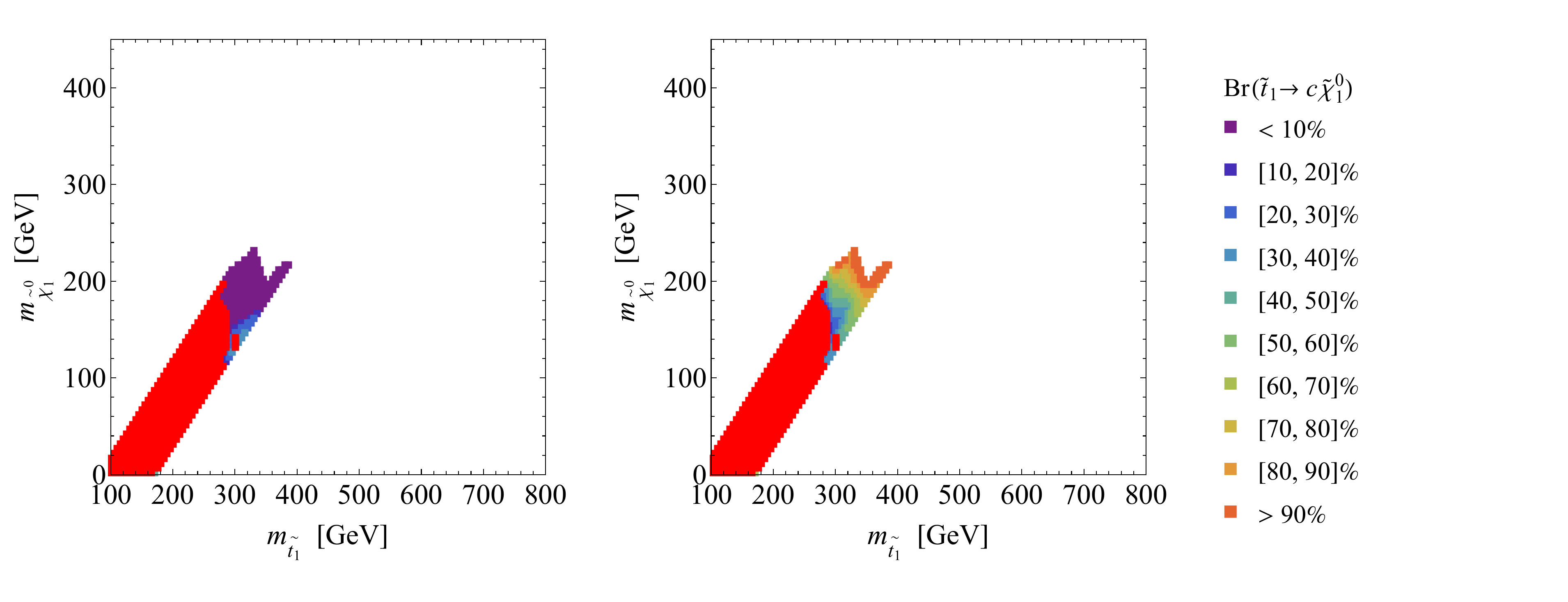} 
\end{center}
\vspace{-8mm}
\caption{\label{fig:exclusionsR2} Model-independent  lower (left panel) and upper (right panel) limits on ${\rm Br} \left (\tilde t_1 \to c \tilde \chi_1^0 \right)$ corresponding to $m_{\tilde{t}_1}$  and $m_{\tilde\chi^0_1}$ values in the kinematic region R2. The regions that are excluded at~95\%~CL for any value of the $\tilde t_1 \to c \tilde \chi_1^0$ branching ratio are coloured red.}
\end{figure}

In Figure~\ref{fig:exclusionsR2}, we show the results of our combination procedure in the case of the kinematic region R2. The parameter space that is excluded at~95\%~CL for any value of the~$\tilde t_1 \to c \tilde \chi_1^0$ branching ratio is shaded red in both panels, while the coloured contours indicate the model-independent limits on ${\rm Br} \left (\tilde t_1 \to c \tilde \chi_1^0 \right)$. We see that $m_{\tilde t_1}$ values up to around~$300 \, {\rm GeV}$ are ruled out by the combined ATLAS Run~I data for essentially all LSP masses $m_{\tilde\chi^0_1}$ satisfying $m_W+m_b<m_{\tilde t_1}-m_{\tilde\chi^0_1}<m_t$. The obtained model-independent bounds on the~$\tilde t_1 \to c \tilde \chi_1^0$ branching ratio will be used in the next section to set limits in the $m_{\tilde{t}_1}$--$\hspace{0.75mm} m_{\tilde\chi^0_1}$ plane for MSSM scenarios with a bino-like LSP and purely right-handed  stop-scharm mixing. 

As in Section~\ref{sec:R1}, we finally discuss the weight  that each search has in our combination. The right panel of Figure~\ref{fig:combinations} shows the 95\% CL exclusion contours in the kinematic region~R2 that are obtained from a naive combination as well as  from a successive inclusion of the searches~$d$, $e$ and $f$. From the figure it is clear that the search~$d$ has by far the strongest effect, and that adding the searches~$e$ and $f$ does not significantly  improve the exclusion in the $m_{\tilde{t}_1}$--$\hspace{0.75mm} m_{\tilde\chi^0_1}$ plane. This again shows that $c$-tagged SUSY searches, like  for instance the~ATLAS analysis \cite{Aad:2014nra}, can also be used to set stringent constraints  on~$m_{\tilde t_1}$ and $m_{\tilde \chi_1^0}$ even outside the kinematic region that the search was initially designed to cover. 

\section{Exclusion limits for purely right-handed up-squark mixing}
\label{sec:RR}

In order to illustrate the effects of flavour mixing in the up-squark sector, we consider a simplified model consisting of a bino-like LSP and a purely right-handed top-like squark $\tilde{t}_1$ that is an admixture of $\tilde{t}_R$ and $\tilde{c}_R$ flavour eigenstates.  Since $\tilde{t}_R\hspace{0.5mm} $--$\hspace{0.25mm}  \tilde{c}_R$ mixing is induced by flavour non-diagonal entries in the squark mass-squared matrix $M_{\tilde u}^2$, it is convenient to use the mass insertion method \cite{Gabbiani:1996hi} to express flavour constraints in terms of the dimensionless quantity
\begin{equation} \label{eq:deltauRRdef}
\delta^u_{RR} = \frac{(M_{\tilde u}^2)_{23}}{ (M_{\tilde u})_{22} \hspace{0.5mm} (M_{\tilde u})_{33}}\,.
\end{equation}
In contrast to quantities like $\delta^{u}_{LL}$ or $\delta^{u}_{RL}$ where mixing with $\tilde{t}_L$ is considered,  purely right-handed scenarios  do not require a light bottom squark, and are hence not subject to strong constraints from direct sbottom searches~\cite{Aad:2013ija,CMS:2014nia,ATLAS-CONF-2015-066,CMS-PAS-SUS-16-004}. Quark flavour constraints are also avoided if only $\delta^u_{RR}$ insertions are considered. We note that although the mass insertion parameter~$\delta^{u}_{LR}$ does not require a light sbottom and is poorly constrained by  flavour physics, its effect on the stop decay width is suppressed relative to $\delta^{u}_{RR}$ by a factor of 16 due to hypercharges. Using  the mass insertion parameter~(\ref{eq:deltauRRdef}) as a template thus allows one to illustrate the maximal effects of flavour violation in  stop decays. Choosing other possibilities would generically lead to stronger exclusions, once additional direct and/or indirect constraints are included.

\begin{figure}
\begin{center}
\includegraphics[height=0.275\textheight]{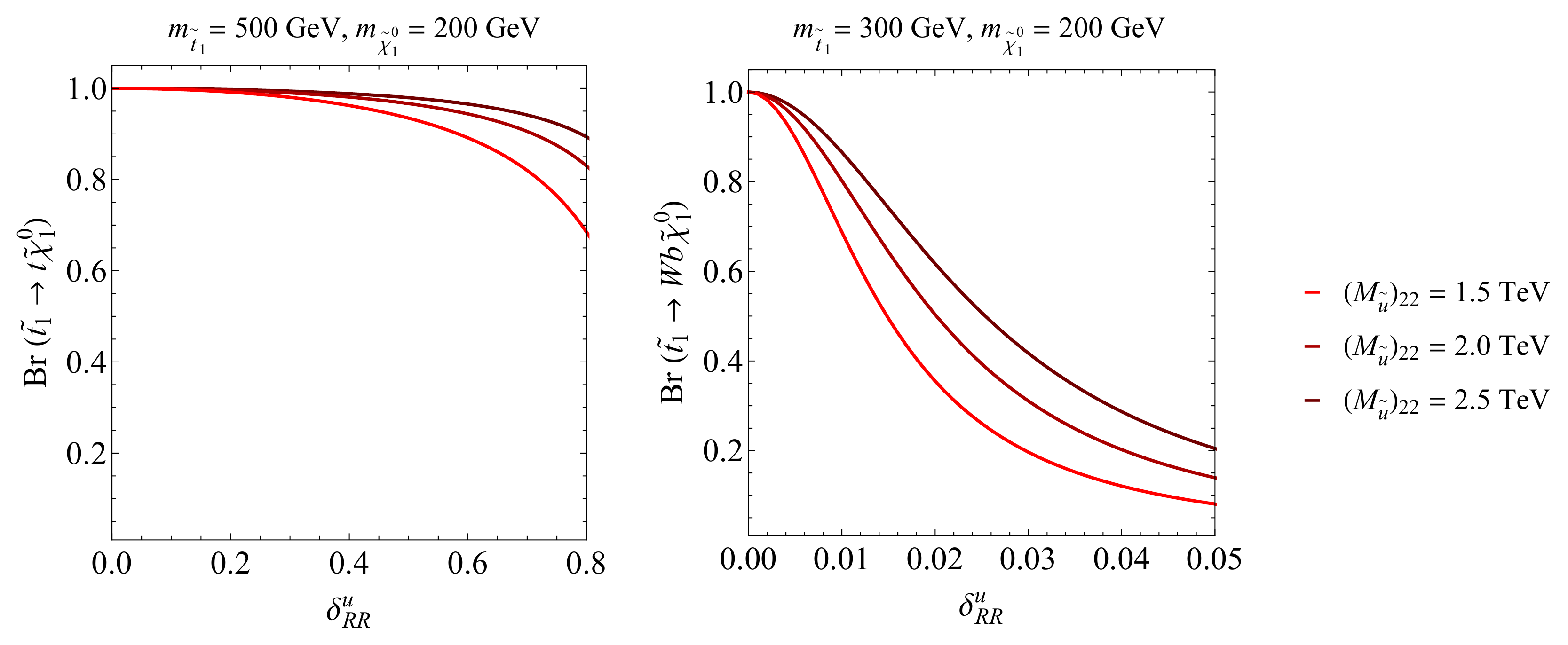} 
\end{center}
\vspace{-8mm}
\caption{\label{fig:deltas} The branching ratios ${\rm Br} \left ( \tilde t_1 \to t \tilde \chi_1^0 \right )$ (left panel) and ${\rm Br} \left ( \tilde t_1 \to Wb \tilde \chi_1^0 \right )$ (right panel) as a function of the mass insertion parameter $\delta_{RR}^u$. The lightest stop mass and the LSP mass have been fixed to $m_{\tilde t_1} = 500 \, {\rm GeV}$ and  $m_{\tilde \chi_1^0} = 200 \, {\rm GeV}$ ($m_{\tilde t_1} = 300 \, {\rm GeV}$ and  $m_{\tilde \chi_1^0} = 200 \, {\rm GeV}$) to obtain  the left (right) plot. The different curves in the panels correspond to different choices of $(M_{\tilde u})_{22}$ as indicated by the legend on the right-hand side in the figure.}
\end{figure}

In Figure~\ref{fig:deltas},  we show the  dependence of ${\rm Br} \left ( \tilde t_1 \to t \tilde \chi_1^0 \right )$ and ${\rm Br} \left ( \tilde t_1 \to Wb \tilde \chi_1^0 \right )$ on $\delta_{RR}^u$ for three different benchmark values of $(M_{\tilde u})_{22}$. The left panel corresponds to $m_{\tilde t_1} = 500 \, {\rm GeV}$ and  $m_{\tilde \chi_1^0} = 200 \, {\rm GeV}$,~i.e.~a parameter point that lies in the heart of the kinematic region~R1.   Similarly, the curves in the  right panel are based on  the point~$m_{\tilde t_1} = 300 \, {\rm GeV}$ and~$m_{\tilde \chi_1^0} = 200 \, {\rm GeV}$, which is located in the kinematic region R2.  Note that even though we parameterise our results in terms of (\ref{eq:deltauRRdef}), we perform an exact diagonalisation in our numerical analysis.  One observes that to reduce ${\rm Br} \left ( \tilde t_1 \to t \tilde \chi_1^0 \right ) = 1 - {\rm Br} \left ( \tilde t_1 \to c \tilde \chi_1^0 \right )$ from~100\% to 90\%  requires  large mass insertions $\delta_{RR}^u \in [0.6, 0.8]$, while  values of~$\delta_{RR}^u \simeq 0.02$ are sufficient to suppress ${\rm Br} \left ( \tilde t_1 \to Wb \tilde \chi_1^0 \right ) = 1 - {\rm Br} \left ( \tilde t_1 \to c \tilde \chi_1^0 \right )$ by approximately~50\%. The strong  $\delta_{RR}^u$-dependence of ${\rm Br} \left ( \tilde t_1 \to Wb \tilde \chi_1^0 \right )$ is  a consequence of the fact that $\tilde t_1 \to Wb \tilde \chi_1^0 $ is a three-body decay, whereas $\tilde t_1 \to c \tilde \chi_1^0$ is a two-body process. In the case of $\tilde t_1 \to t \tilde \chi_1^0$, the flavour-conserving decay mode is not phase-space suppressed and as a result a larger $\tilde{t}_R\hspace{0.5mm} $--$\hspace{0.25mm}  \tilde{c}_R$ mixing angle is needed to obtain appreciable values of~${\rm Br} \left ( \tilde t_1 \to c \tilde \chi_1^0 \right )$.  In the kinematic region~R3,~$\tilde t_1 \to c \tilde \chi_1^0$ is the dominant decay mode unless $\delta_{RR}^u$ is below $10^{-3}$, in which case the four-body mode~$\tilde t_1 \to b f f^\prime \tilde \chi_1^0$ becomes the main channel. In the following, we will not consider such small $\tilde{t}_R\hspace{0.5mm} $--$\hspace{0.25mm}  \tilde{c}_R$ mixing angles and hence will employ~${\rm Br} \left ( \tilde t_1 \to c \tilde \chi_1^0 \right ) = 100\%$ in the whole region R3.

\begin{figure}
\begin{center}
\includegraphics[height=0.3\textheight]{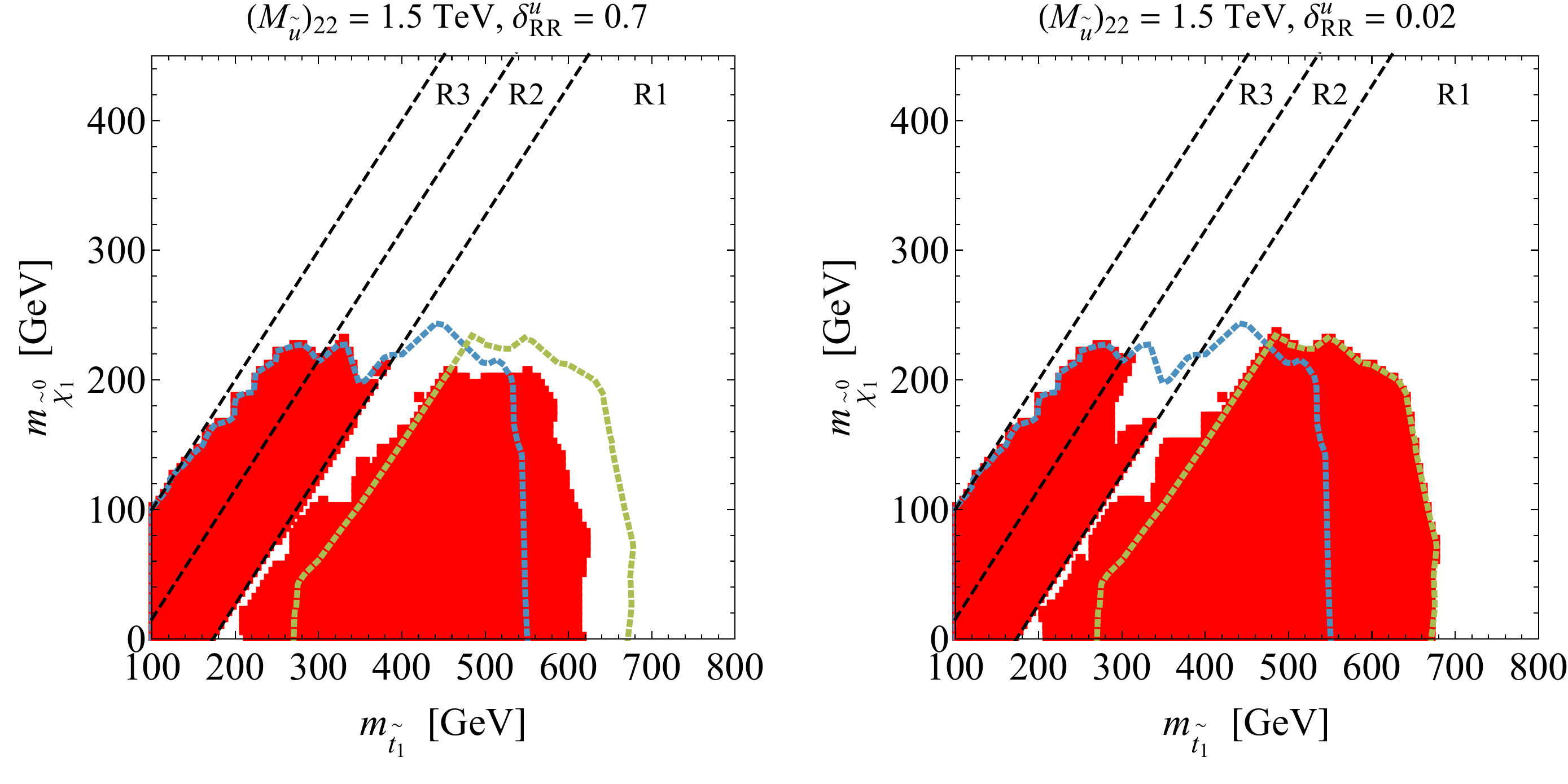} 
\end{center}
\vspace{-8mm}
\caption{\label{fig:RRscenarios} 95\%~CL exclusion regions in the  $m_{\tilde{t}_1}$--$\hspace{0.75mm} m_{\tilde\chi^0_1}$  plane for two representative $\tilde{t}_R\hspace{0.5mm} $--$\hspace{0.25mm}  \tilde{c}_R$ mixing scenarios. The left (right) panel employs the parameters $(M_{\tilde u})_{22} = 1.5 \, {\rm TeV}$ and $\delta_{RR}^u = 0.7$ $\big($$(M_{\tilde u})_{22} = 1.5 \, {\rm TeV}$ and $\delta_{RR}^u = 0.02$$\big)$. For comparison also the exclusion limits at 95\%~CL following from the $1~\text{lepton} + 4~\text{jets} + 1~b\text{-tag} + E_{T, \rm miss}$ search \cite{Aad:2014kra} (green dotted curves) and the $2~c\text{-tags} + E_{T, \rm miss}$ search~\cite{Aad:2015gna} (blue dotted curves) are overlaid.}
\end{figure}

In the two panels of Figure~\ref{fig:RRscenarios}, we show the results of the combined search strategies of Sections~\ref{sec:R1} and \ref{sec:R2} when applied to two representative scenarios of~$\tilde{t}_R\hspace{0.5mm} $--$\hspace{0.25mm}  \tilde{c}_R$ mixing. The left panel depicts the case of large mixing $\delta_{RR}^u = 0.7$, while the right panel illustrates the case  of small mixing $\delta_{RR}^u = 0.02$. In both plots, we have fixed $(M_{\tilde u})_{22} = 1.5 \, {\rm TeV}$ and the red contours correspond to the regions in the $m_{\tilde{t}_1}$--$\hspace{0.75mm} m_{\tilde\chi^0_1}$  plane that are excluded at~95\%~CL. To guide the eye, the 95\%~CL exclusion limits obtained in \cite{Aad:2014kra} and~\cite{Aad:2015gna} have been overlaid as green and blue dotted curves. Focusing our attention on the kinematic region~R1, we see that for the choice $\delta_{RR}^u = 0.7$ the  limits on $m_{\tilde t_1}$ are about~$50 \, {\rm GeV}$ weaker than the  bounds obtained in \cite{Aad:2014kra}, which assumes no stop-scharm mixing. On the other hand,  for~$\delta_{RR}^u = 0.02$  our exclusion coincides with the limit of the $1~\text{lepton} + 4~\text{jets} + 1~b\text{-tag} + E_{T, \rm miss}$ search. These features are expected because in the first case one has ${\rm Br} \left ( \tilde t_1 \to t \tilde \chi_1^0 \right ) \in [70, 80]\%$ in the parameter space of interest, while ${\rm Br} \left ( \tilde t_1 \to t \tilde \chi_1^0 \right ) \simeq 100 \%$ in the second case. In the kinematic region R2, one observes instead that for large  $\tilde{t}_R\hspace{0.5mm} $--$\hspace{0.25mm}  \tilde{c}_R$ mixing our bound resembles that of the analysis~\cite{Aad:2015gna}, while for small mixing the region in the $m_{\tilde{t}_1}$--$\hspace{0.75mm} m_{\tilde\chi^0_1}$  plane around~$m_{\tilde t_1} = 300 \, {\rm GeV}$ and $m_{\tilde \chi_1^0} = 200 \, {\rm GeV}$ remains allowed.  These properties can be understood by realising  that in the first case the lightest stop decays to almost $100 \%$ via~$\tilde t_1 \to c \tilde \chi_1^0$, while in the second case the decay mode $\tilde t_1 \to Wb\tilde \chi_1^0$ is dominant, in particular for values of the stop and LSP mass close to the kinematic boundary $m_{\tilde t_1} - m_{\tilde \chi_1^0} = m_t$. One furthermore notices, that in region R3 our exclusions  match the 95\% CL bound from the~$\tilde c_1 \to c \tilde \chi_1^0$ analysis~\cite{Aad:2015gna}, since both our choices of $\delta_{RR}^u$ lead to ${\rm Br} \left ( \tilde t_1 \to c \tilde \chi_1^0 \right ) = 100 \%$. The two scenarios of $\tilde{t}_R\hspace{0.5mm} $--$\hspace{0.25mm}  \tilde{c}_R$ mixing that we have considered nicely illustrate our general finding that by combining various $E_{T, \rm miss}$ search strategies, large regions in the $m_{\tilde{t}_1}$--$\hspace{0.75mm} m_{\tilde\chi^0_1}$  plane can be excluded for arbitrary mass insertion parameters $\delta_{RR}^u$. 

Notice that quark flavour observables leave the mass insertion parameter $\delta^{u}_{RR}$ essentially unconstrained. Although~$\tilde{t}_R\hspace{0.5mm} $--$\hspace{0.25mm}  \tilde{c}_R$ mixing will induce flavour-changing top-quark decays like $t \to cZ$ and $t \to ch$ at the one-loop level, the existing LHC Run I constraints on  the relevant processes \cite{Loginov:2015nxw} are too loose to lead to any restriction.  The mass insertion parameter~$\delta^{u}_{RR}$ also modifies $B$-meson decays via chargino loops.  However, the wino couples only to left-handed squarks and the Higgsino coupling to right-handed squarks is proportional to the corresponding Yukawa coupling, which is small in the case of the charm squark.  As a result, corrections associated to the $\delta^{u}_{RR}$ mass insertion are strongly suppressed in processes like $B_s \to \mu^+ \mu^-$ and $B \to X_s \gamma$, making the constraints from stop  searches derived above the only relevant restrictions on scenarios with purely right-handed  stop-scharm mixing. 

\section{Conclusions and outlook}
\label{sec:conclusions}

In this article, we have shown that allowing for non-minimal flavour violation in the up-squark sector of the MSSM can weaken the direct LHC bounds on the mass~$m_{\tilde{t}_1}$ of the lightest stop. While  large effects were found previously~\cite{Bartl:2012tx,Blanke:2013zxo,Agrawal:2013kha}, we have demonstrated that a detailed numerical analysis  which includes  the recent ATLAS search for $\tilde c_1\to c\tilde\chi^0_1$ limits the possible impact of $\tilde t_1\to c\tilde\chi^0_1$ on $\tilde t_1\to c \tilde\chi^0_1$ and $\tilde t_1\to Wb \tilde\chi^0_1$. The general idea is that although an enhanced $\tilde t_1\to c\tilde\chi^0_1$ decay rate  decreases the branching ratios of $\tilde t_1\to t \tilde\chi^0_1$ and~$\tilde t_1\to Wb \tilde\chi^0_1$, the direct $\tilde c_1\to c\tilde\chi^0_1$ bounds become progressively more relevant, and as a result stop and scharm searches cannot be fully decoupled in the presence of up-squark mixing.  By combining the different decay channels, we demonstrated that there are large regions in the $m_{\tilde{t}_1}$--$\hspace{0.75mm} m_{\tilde\chi^0_1}$ plane which are disfavoured by LHC~Run~I searches, independently of the amount of stop-scharm mixing. In particular, we find a lower limit of $m_{\tilde t_1} > 530 \, {\rm GeV}$ at 95\%~CL for LSP masses $m_{\tilde \chi^0_1} \lesssim 100 \, {\rm GeV}$. This finding agrees with \cite{Blanke:2015ulx}, generalising it to the case of a neutralino with non-zero mass. Stringent exclusion limits can also be derived for all other considered decay scenarios.  We have illustrated this point by studying MSSM scenarios with a bino-like LSP and  non-zero~$\tilde{t}_R\hspace{0.5mm} $--$\hspace{0.25mm}  \tilde{c}_R$ mixing. The two representative cases of the  mass insertion parameter $\delta_{RR}^u$ that we have considered are left unconstrained by quark flavour observables, but by combining various direct  $E_{T, \rm miss}$ searches, stringent exclusions in the $m_{\tilde{t}_1}$--$\hspace{0.75mm} m_{\tilde\chi^0_1}$ plane can be derived. 

In LHC Run~II and beyond, the ATLAS and CMS collaborations are expected to provide new results on stop searches with a siginificantly improved reach in the  $m_{\tilde{t}_1}$--$\hspace{0.75mm} m_{\tilde\chi^0_1}$ plane.  Improvements in the sensitivity  to stops will not only be due to the increase in the centre-of-mass energy, but is also likely to arise from new analysis strategies or  technical developments. For instance, ATLAS has recently installed~\cite{Phoboo:1998718} a new subdetector  called Insertable B-Layer or IBL~\cite{Capeans:1291633}. This new inner pixel layer should allow to improve  the $c$-tagging capabilities of ATLAS and thus pave the way  to look for processes like $\tilde c_1 \to c \tilde \chi_1^0$  and $\tilde t_1 \to c \tilde \chi_1^0$ in a more efficient fashion.  The complementarity and synergy between  the different stop decay channels that exists in the presence of flavour mixing is therefore expected to become phenomenologically even more relevant at later phases of the LHC physics programme. 

\acknowledgments

We thank William Kalderon for useful discussions concerning the ATLAS search for scalar charm quarks~\cite{Aad:2015gna}. The work of AC was supported by a Marie Curie Intra-European Fellowship of the European Community's 7th Framework Programme (contract number PIEF-GA-2012-326948) and by an Ambizione Grant of the Swiss National Science Foundation. UH acknowledges the hospitality and support of the CERN theory division. He also would like to thank the KITP in Santa Barbara for hospitality and acknowledges that this research was supported in part by the National Science Foundation under Grant No.~NSF~PHY11-25915. LCT is supported by the Swiss National Science Foundation. 

\begin{appendix}

\section{Event generation}
\label{app:A}

Our event generation has been performed at leading order with {\tt MadGraph5\_aMC\@NLO}~\cite{Alwall:2014hca} starting from a customised version of the implementation of coloured scalar pair production presented in~\cite{Degrande:2014sta} and utilises {\tt NNPDF2.3} parton distribution functions~\cite{Ball:2012cx}. The simulated parton-level events were showered with  {\tt  PYTHIA~6}~\cite{Sjostrand:2006za} and analysed with the publicly available code {\tt CheckMATE}~\cite{Drees:2013wra}, which relies on {\tt DELPHES~3}~\cite{deFavereau:2013fsa} as a fast detector simulation.  In order to be able to distinguish charm-quark jets from both bottom-quark and light-flavoured jets, we have implemented the {\tt JetFitterCharm} algorithm described in~\cite{ATL-PHYS-PUB-2015-001} into {\tt DELPHES~3}. In all our analyses jets were clustered with {\tt FastJet}~\cite{Cacciari:2011ma} with the anti-$k_t$ algorithm~\cite{Cacciari:2008gp} as the standard jet finder. 

The efficiency maps presented in Figures~\ref{fig:efficiencies1} and~\ref{fig:efficiencies2}  have been obtained by simulating~138 different signal points that fall into the kinematic regions R1 and R2. The actual mapping  in the  $m_{\tilde{t}_1}$--$\hspace{0.75mm} m_{\tilde\chi^0_1}$ plane  can be found in \cite{AAD2014HepData, AAD2015HepData}. For each signal point and all relevant final states, $10^5$ partonic events have been generated, showered and passed through  the fast detector simulation and an analysis containing the selection requirements corresponding to the individual searches described in Sections~\ref{sec:R1} and \ref{sec:R2}. The efficiency maps are then obtained by considering the SR that provides the best exclusion limit for a given point in the  $m_{\tilde{t}_1}$--$\hspace{0.75mm} m_{\tilde\chi^0_1}$ plane.

\end{appendix}

\providecommand{\href}[2]{#2}\begingroup\raggedright\endgroup

\end{document}